\documentclass{appolb}
\usepackage{epsfig}
\usepackage{psfig}

\begin{document}
\eqsec  
\title{LHC-ILC synergy
\thanks{Presented at PLC2005, Kazimierz, September 5-8, 2005.}%
}
\author{Rohini M. Godbole
\address{Center for High Energy Physics, Indian Institute of Science, Bangalore,560012, India.}
}
\maketitle
\begin{abstract}
I will begin by making a few general comments on the synergy between the Large
Hadron Collider (LHC) which will go in action in 2007 and the International 
Linear Collider (ILC) which is under planning. I will then focus on the synergy between the LHC and the PLC option at the ILC, which is expected to be realised
in the later stages of the ILC program. In this I will cover the possible 
synergy in the Higgs sector (with and without CP violation), in the 
determination of the anomalous vector boson couplings and last but not the 
least, in the search for extra dimensions and radions.
\end{abstract}
  
\section{Introduction}
Historically there has always been  feedback and interplay between the 
hadronic and the leptonic colliders. $Sp\bar pS$ saw a handful of $W'$s and 
$Z'$s, establishing the correctness of the $SU(2) \times U(1)$ model, 
whereas the LEP/SLC tested it to a one per mil precision using the millions of 
$Z'$s and thousands of $W's$. The agreement between the `prediction' of the 
top mass obtained using precision measurements from  LEP and the `direct' 
measurements made at the Tevatron, was indeed a very important step in 
establishing the Standard Model. But since this synergy has always existed,
one may well ask the question as to what is the special need NOW for 
discussing the LHC/ILC synergy. The need arises  from the current state of 
play in High Energy Physics (HEP) and the high stakes in physics studies at 
future colliders, both 
on the physics front and on the economic front; as well as the long time 
scales which the  planning and execution for a new collider require.  
\begin{figure}[htb]
\includegraphics*[scale=0.8]{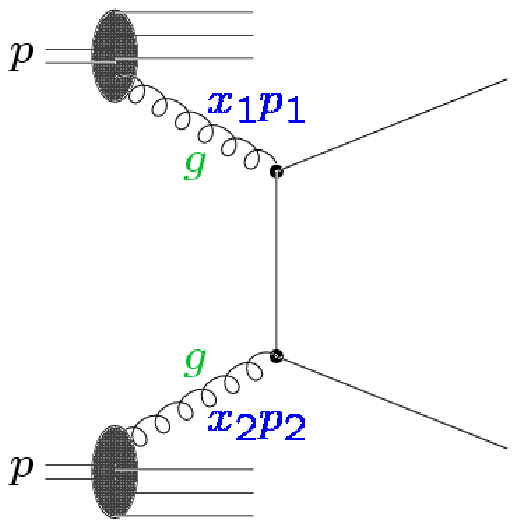}
\includegraphics*[scale=0.8]{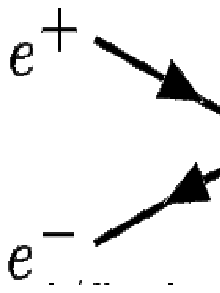}
\caption{The LHC and the ILC.  \label{fig1}}
\end{figure}
The LHC is a hadronic collider with pp collisions at  $\sqrt{s}  = 14$ TeV,
whose strong point is the larger mass reach for direct discoveries. Even 
though 
at the LHC the composite nature of colliding protons gives rise to an 
underlying event and the $\sqrt{s}$ of the hard interaction is not fixed, one 
can use conservation of the transverse momentum $P_T$ and thus study 
interesting 'hard' physics. Being a hadronic machine all the physics studies
at the LHC will have to deal with large QCD backgrounds. The ILC on the other 
hand is planned to be an  $e^+e^-$ collider with  $\sqrt{s}  = 0.5$ -- $1.0$
TeV. Its strong point being the  high precision  physics. The precisely known 
initial state kinematics along with possibilities of the initial state beam 
polarisation,  allows accurate and detailed analysis of the decays, precision  
determinations of masses and couplings etc. Since the initial state contains 
EW particles the QCD  backgrounds will be smaller than at the LHC. Further, 
it will also offer  possibilities of various options; such as the 
$\gamma \gamma$, $\gamma e$ and $e^- e^-$, opening up chance to study
aspects of physics of the SM as well as physics beyond the SM, that may not
accessible at the LHC or in the $e^+e^-$ option of the ILC.  The LHC, however,
has the greatest  advantage of all, viz.,  it is all geared  to start action in 
2007, whereas we still are not sure IF, WHEN and WHERE construction will 
happen for the ILC.  It is heartening however that a clear international 
consensus has emerged on the ILC now and the planning is in full swing.
\begin{figure}[htb]
\includegraphics*[scale=0.35]{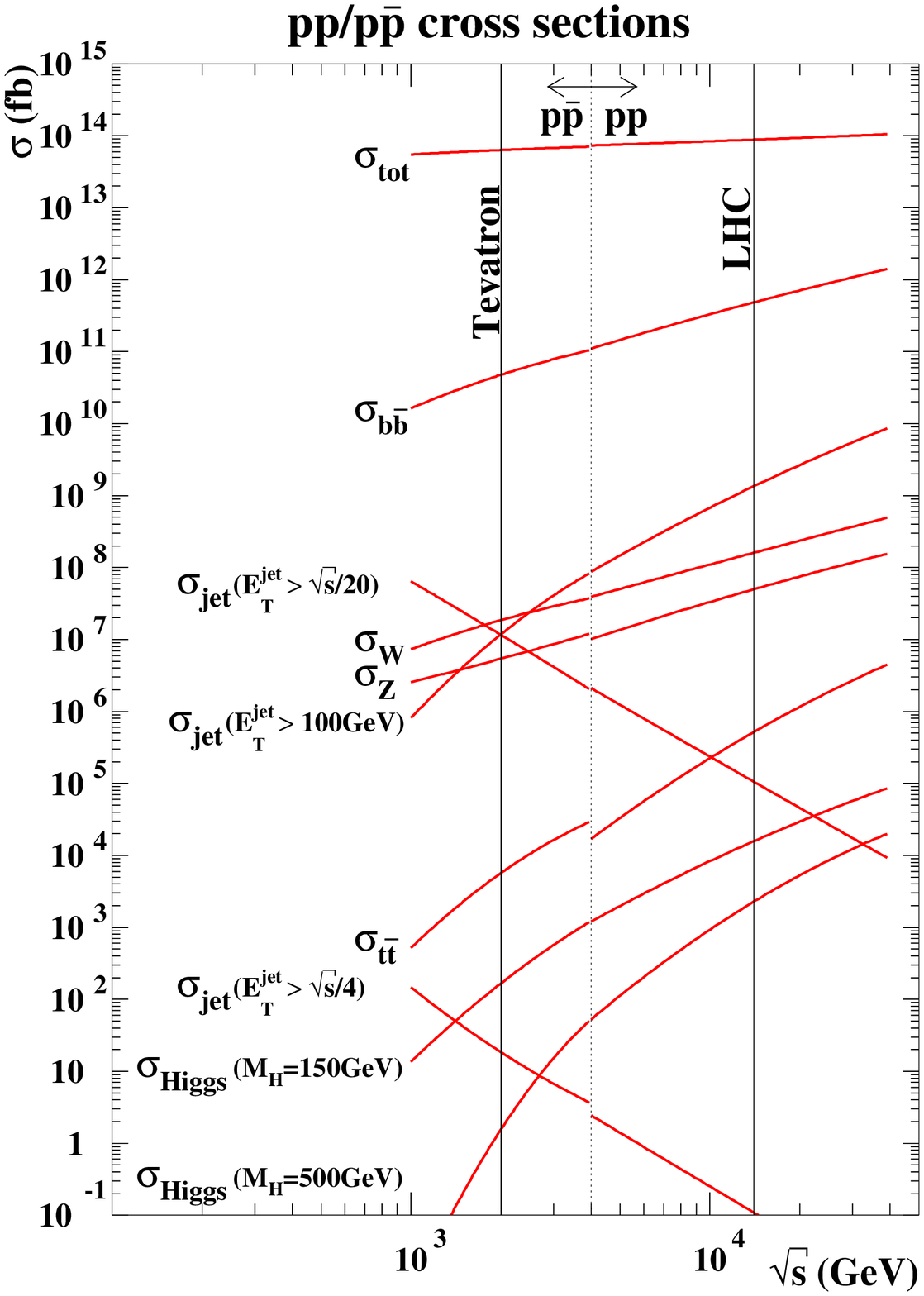}
\includegraphics*[scale=0.35]{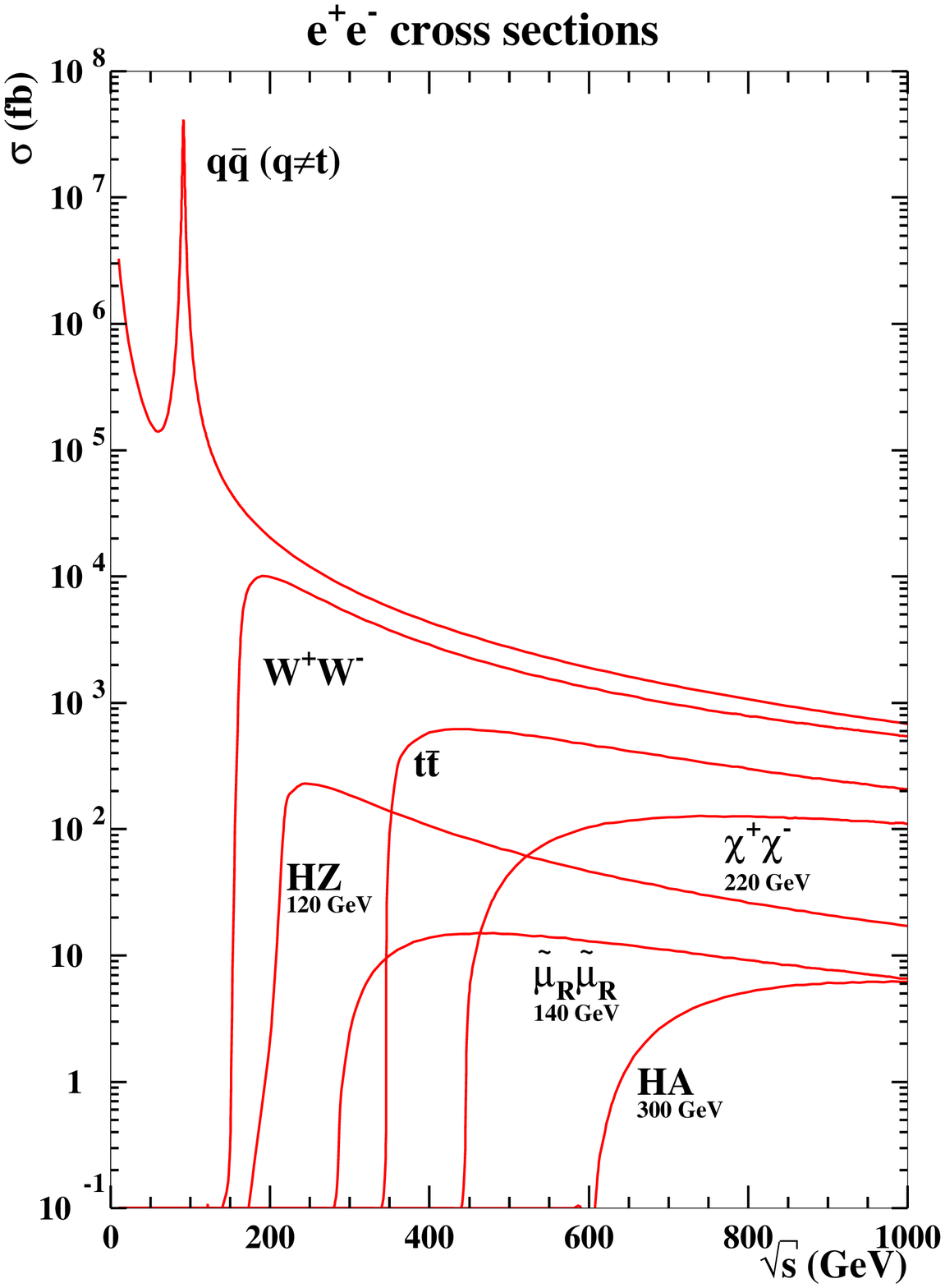}
\caption{Expected cross-sections for the physics process of interest at the
LHC and the ILC\protect\cite{lhc-ilc}.\label{fig2}}
\end{figure}

Fig.~\ref{fig2} shows the expected cross-sections at the LHC and the ILC for 
different processes.  The plots in left panel show us, for example,
that in the exploration  
of the Higgs physics which would be the focus of studies at the LHC, it is 
going to be a big challenge to deal with the large  physics backgrounds. 
On the other hand we see from the expected cross-sections at the ILC for the 
different physics processes, that it is the ILC which will offer the 
possibility of high precision study of the Higgs sector. Therefore,  as 
a community it is very important for us to assess the desired energy, 
luminosity and {\bf the timing} of the ILC vis-a-vis the physics goals of the 
LHC/ILC which are set by the current stage of understanding in particle 
physics. In the context of the PLC it really means making sure that the 
ILC designs keep the possibilities of the PLC option open. 

\section{LHC-ILC interplay}
Till about 2002 or so there was not much interaction between the LHC and the LC 
community.  LHC/ILC study group was first formed in the context of an ECFA 
LC group and then became a worldwide affair. It has been a collaborative effort 
of the Hadron Collider (the LHC) and the Leptonic Linear Collider (ILC) 
community. At the time of the formation of the group the LHC was well on its 
way and the physics case of an ILC had been clearly made. (Incidentally 
for the PLC this exercise 
needs to be cemented beyond the studies that have already been 
made~\cite{tesla-plc}.) The aim of the LHC/ILC study group was NOT to 
compare which of the two colliders can do better but rather how the two 
can complement each other and further whether one can identify areas  where 
the cross-talk between the two colliders can increase the utility of {\it both}.It is clear that LHC will  have higher reach in energy and hence can perhaps 
create directly new particles expected in the extensions of the SM.
The ILC on the other hand can make precision measurements and can be sensitive 
to the indirect effects of the same particles even if  masses are much higher
than the energy of the ILC.  Thus information from a lower energy ILC can still 
feedback into studies at the  LHC. This is the simplest form of synergy between
the two colliders. We have seen an example of this in the  comparison 
between the mass of the top quark as 
estimated from the precision EW  measurements and as measured directly from 
the Tevatron data. Now we see similar interplay for the prediction of  the
(SM) Higgs mass, being sharpened by knowledge of the top mass from Tevatron.
Precision measurements from the ILC may  therefore sometimes be able to tell
the LHC where to focus the effort. Precision measurements at the LHC are 
difficult if not impossible, but will be possible only after a few years after 
the beginning of its operation.  These studies can definitely benefit due to 
the feedback from the ILC. Of course the ability of the ILC is not restricted 
to precision measurements alone but also to making possible discoveries which 
at times will be difficult or impossible at the LHC.  These qualitative 
statements are almost obvious to practicing phenomenologists and 
experimentalists,  but  quantitative studies are necessary. Various examples of 
such studies and possible cross-talks are  present in the 
document~\cite{lhc-ilc}.

\noindent 
In the study group report~\cite{lhc-ilc} possibilities were analysed assuming 
that the LHC will run for 20 years and that the ILC can kick off after the 
LHC has been running a few years.  
The specific questions that were addressed in this document were as follows:
\begin{itemize}
\item[1] 
How information obtained at both the colliders can be put {\bf together} so 
that the basic physics questions being asked by the HEP community can be 
answered more {\bf conclusively and effectively}.
\item[2]
Can the combined studies  give pointers to new bench marks for measurements 
at the LHC.
\item[3] Can the results obtained at a lower energy ILC affect the analysis 
at the LHC if {\bf not} the triggering. Can it  affect the luminosity/detector 
upgrades and also provide yet more focus to the LHC studies.
\item[4]
What are the physics needs and advantages of concurrent running of the LHC 
and the ILC.

\item[5] What are the physics arguments to make  a strong case  for 
keeping the door for PLC open in the ILC designs under consideration.
\end{itemize}

One can think of  various possible scenarios for the cross-talk:
\begin{itemize}
\item[1)]LHC + ILC :
ILC data help clear up the underlying structure of new physics of 
which Tevatron and the LHC will give us some glimpses.

\item[2)]
{\parbox{4.5cm}{\includegraphics*[scale=0.4]{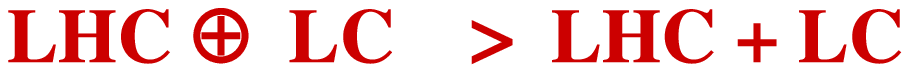}}}\hspace{-.5cm}:
A combined interpretation of LHC/ILC data  can help use both the data
more effectively to learn about the TeV scale physics beyond the SM;
in particular such an analysis can reduce possible model dependencies.

\item[3)]
\parbox{4.5cm}{\includegraphics*[scale=0.4]{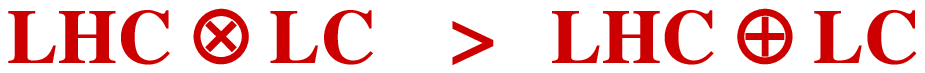}}\hspace{-0.5cm}:
If the machines have some overlap in time, and a combined analysis of the 
LHC/ILC data were to be possible then the  ILC results could influence the 
second phase of the LHC, just like some of the LEP-II results have affected
the Tevatron upgrades.  Similarly, the ILC results could  provide inputs to 
the upgrade options for the LHC machines and detectors.
\end{itemize}

A few comments are in order here:
While no examples could be found such that the 'triggering' at the LHC could 
be affected by what the ILC will see, a very good case could be made for aiming
at a combined interpretation of the data and some time overlap so that ILC 
affect the upgrades at LHC.  Many examples for the latter were 
found particularly in the context of SUSY studies. In the context of analysis 
in the Higgs sector, it was shown that a reduction of model dependencies may 
be achieved through a combined LHC/ILC analysis. We will look at one example 
each from the Higgs and the SUSY sector and then go over to the case of a PLC.

\subsection{Higgs studies} 
The LHC will be able to observe the SM Higgs and afford measurements of 
its various properties such as width, relative couplings to some accuracy (about
15-20\%) by the end of the high luminosity run~\cite{atlas-tdr,abdel-hak}. As 
far as the ILC is concerned it can of course profile a Higgs most accurately 
even in the low energy, 
moderate luminosity option~\cite{tesla-tdr}, except for the measurement of the
$t \bar t H$  coupling and the reconstruction of the Higgs potential.  
At the ILC, a  precision measurement of $t \bar t H$  coupling requires 
$\sqrt{s} = 800$ -– $1000$ GeV.  The LHC measures $\sigma \times$ B.R. 
into  different  channels.  One question that can be asked is whether a cross 
talk between the LHC and the ILC can improve this situation?
\begin{figure}
\includegraphics*[scale=0.6]{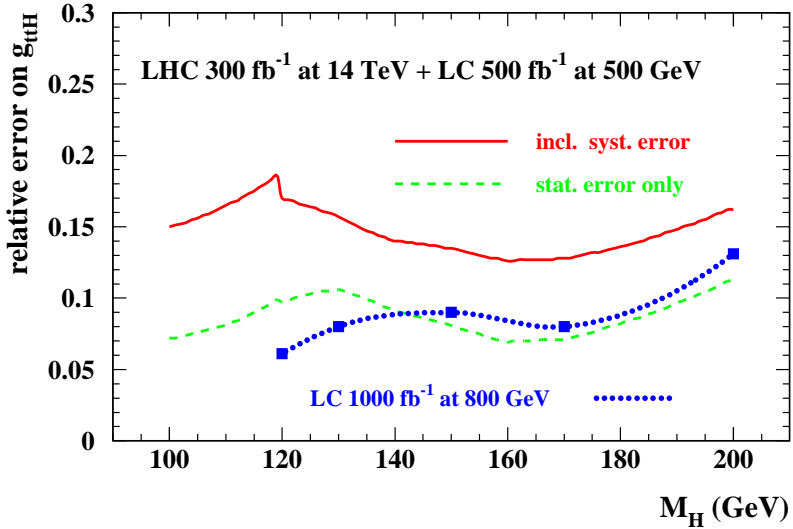}
\caption{LHC-ILC cross talk for the $t \bar t$ Higgs coupling determination\protect\cite{higgs-lhc-ilc}.
\label{fig3}}
\end{figure}
The strategy  is to use the ILC precision information  on the other 
branching ratios of the Higgs and thus get information on the 
$t \bar t H$  coupling in a model  independent way, using BOTH  the ILC 
and the LHC data. As can be seen from Fig.~\ref{fig3} the combined use 
of the LHC and a 500 GeV ILC with $500$ fb$^{-1}$ integrated luminosity, 
will allow a determination of the $t \bar t H$ coupling to an accuracy 
of $10\%$--$15\%$. Further, the accuracy expected taking into account 
the statistical errors alone is about $5\%$ which is comparable to that 
expected for a 800 GeV ILC with $1000$ fb $^{-1}$. This clearly shows 
that the combined analysis of the LHC and the ILC data gives 
better value for  money~\cite{higgs-lhc-ilc}.

\subsection{Supersymmetry studies} 
\begin{figure}
\begin{center}
\includegraphics*[scale=.4]{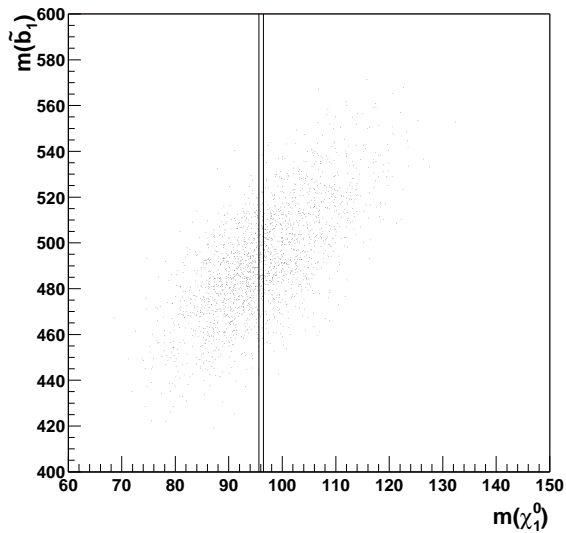}
\end{center}
\caption{Mass correlation plots, where the dots show possibilities using the 
LHC result alone and the vertical bands the precision possible on the LSP mass 
determination at the LSP\cite{susy-lhc-ilc}.\label{fig4}}
\end{figure}

Next to Higgs searches and study of its properties, Supersymmetry (SUSY) 
searches~\cite{SUSY_book} will  form an important part of the physics program 
of any collider, be it hadronic, leptonic or photonic.
Supersymmetry is certainly broken since  we do not  see the superpartners of 
the particles of the SM,  differing in spin by $1/2$ from them and with the
same mass as them. In the unconstrained minimal supersymmetry standard model,
MSSM, there exist 105 new parameters in the form of the masses of the 
superpartners and mixing angles.  Normally while investigating the 
prospect of SUSY studies at the LHC, one reduces the number of these parameters
by working in the framework of one of the SUSY breaking models. These are 
named after the mechanism used for SUSY breaking. These are: 
a) Gravity mediated (MSUGRA), b) Gauge mediated SUSY breaking, 
c) Anomaly mediated SUSY breaking etc. If TeV scale SUSY should exist then 
the probability that the LHC will see some signal is very high. The current 
studies of the LHC potential in the context of SUSY not only look
at the prospects of discovering it but focus also on  the possible
measurements of masses and mixing angles. These in turn  may be used for
SUSY parameter and consequently the SUSY breaking mechanism determination. 
So far most of the studies of the LHC potential have been model dependent, 
now they move to  {\bf model independent} ones. At the LHC masses will not 
determined  with very high  precision  but mass differences will be. In the 
commonly used 
$R$--parity conserving SUSY scenarios, all the final states corresponding 
to a decay chain of a given sparticle contains at least one lightest 
supersymmetry particle (LSP) which appears as 'missing' energy. Thus the 
sparticle mass determined is highly correlated with the mass of the LSP.
This is illustrated in Fig.~\ref{fig4} for the determination of the 
$\tilde b_1$ mass.
The dots in Fig.~\ref{fig4} indicate the results achievable using the LHC data 
alone.  On the other hand, at the ILC an accurate determination of
$m_{\tilde{\chi}_1^0}$ is possible. The vertical bands in the Fig.~\ref{fig4} 
shows the results if one were to restrict the $m_{\tilde{\chi}_1^0}$ to within
$\pm2\sigma$ with the ILC input ($\sigma=0.2\%$). So the suggested strategy is 
to use the accurate mass determination of $\tilde \chi_1$ from
the ILC and feed it in LHC sparticle mass determination. The second column in 
the Table 1 shows that indeed the ILC input can reduce the errors in the 
sparticle mass determination substantially. These investigations brought out
an interesting feature, that the jet measurement seems to be the limiting 
factor for the accuracies possible with a combined analysis of the LHC and the
ILC data. {\it This is an example where the study has isolated a feature of
the LHC analysis which, if improved upon, can add to  the accuracy of the 
sparticle mass determination at the LHC in a big way.} This thus is a very
good example of the LHC-ILC synergy.
\begin{table}
\begin{tabular}{cccc}
\hline\hline
 & LHC & LHC+ILC  \\
\hline
$\Delta m_{\tilde{\chi}_1^0}$& 4.8 & 0.05 (ILC input) \\
$\Delta m_{\tilde{\chi}_2^0}$& 4.7 & 0.08 \\
$\Delta m_{\tilde{\chi}_4^0}$ & 5.1 & 2.23 \\
$\Delta m_{\tilde{l}_R}$ & 4.8 & 0.05 (ILC input)  \\
$\Delta m_{\tilde{l}_L}$ & 5.0 & 0.2 (ILC input) \\
$\Delta m_{\tilde{\tau}_1}$ & 5-8 & 0.3 (ILC input) \\
$\Delta m_{\tilde{q}_L}$ & 8.7  & 4.9  \\
$\Delta m_{\tilde{q}_R}$ & 7-12 & 5-11  \\
$\Delta m_{\tilde{b}_1}$ & 7.5 & 5.7  \\
$\Delta m_{\tilde{b}_2}$ & 7.9 & 6.2  \\
$\Delta m_{\tilde{g}}$ & 8.0 & 6.5  \\
\hline
&&\\
\end{tabular}
\caption{
The improvement in the possible precision in the sparticle mass 
determination due to the combined use of the LHC and the ILC data. The errors
quoted are in GeV and are for the point SPS1a~\protect\cite{susy-lhc-ilc}.
\label{table1}}
\end{table}

\section{PLC and LHC/ILC($e^+e^-$) synergy}
\subsection{Higgs and SUSY}
To begin with the accurate ($\sim 2 \%$)  measurement of the 
$\gamma \gamma$ decay width of a light Higgs boson possible at the 
PLC~\cite{tesla-plc}, allows a probe of high scale physics as the heavy 
particles affect this decay width through loop effects. The  
availability of polarized photon spectra and a democratic mechanism for 
production of CP-even and CP-odd Higgs, makes the PLC an ideal tool to
probe CP-violation in the Higgs sector. Further, the $s$-channel production
mechanism allows for single-Higgs production and hence increases the reach
compared to the $e^+e^-$ option by about a factor of $1.6$. As a matter of
fact, in the  MSSM for $\tan \beta \simeq 4-10$, $M_A, M_H > 200$--$250$ 
GeV, the LHC will see only one spin 0 state and the $H,A$ are not accessible 
for the first generation, 500 GeV, ILC. The PLC offers possibilities of
probing the $H/A$ in this  so called 'LHC wedge' 
region~\cite{maggie2,maria2,maria1,maria3} through their
$s$--channel production and decay into a $b \bar b$ and $WW/ZZ$ final 
states. For larger values of $\tan \beta$, where the $b \bar b$ final state
can not be used effectively, the decays into a neutralino pairs can be used 
too~\cite{maggie2}. The PLC also offers a possibility of pinning down
the Higgs structure of a theory in a general 2 Higgd doublet model (2HDM).
Some aspects of this have already been discussed elsewhere in the 
proceedings~\cite{others,mesusy}. 

Krawczyk and collaborators~\cite{lhc-ilc-2hdm} have discussed and example where
the measurement of $\gamma \gamma$ width of the Higgs is essential to determine
whether a Higgs seen at the LHC is indeed a SM Higgs. These authors have 
identified realisations of the 2HDM with a SM-like light Higgs boson.
\begin{figure}[htb]
\includegraphics*[scale=0.22]{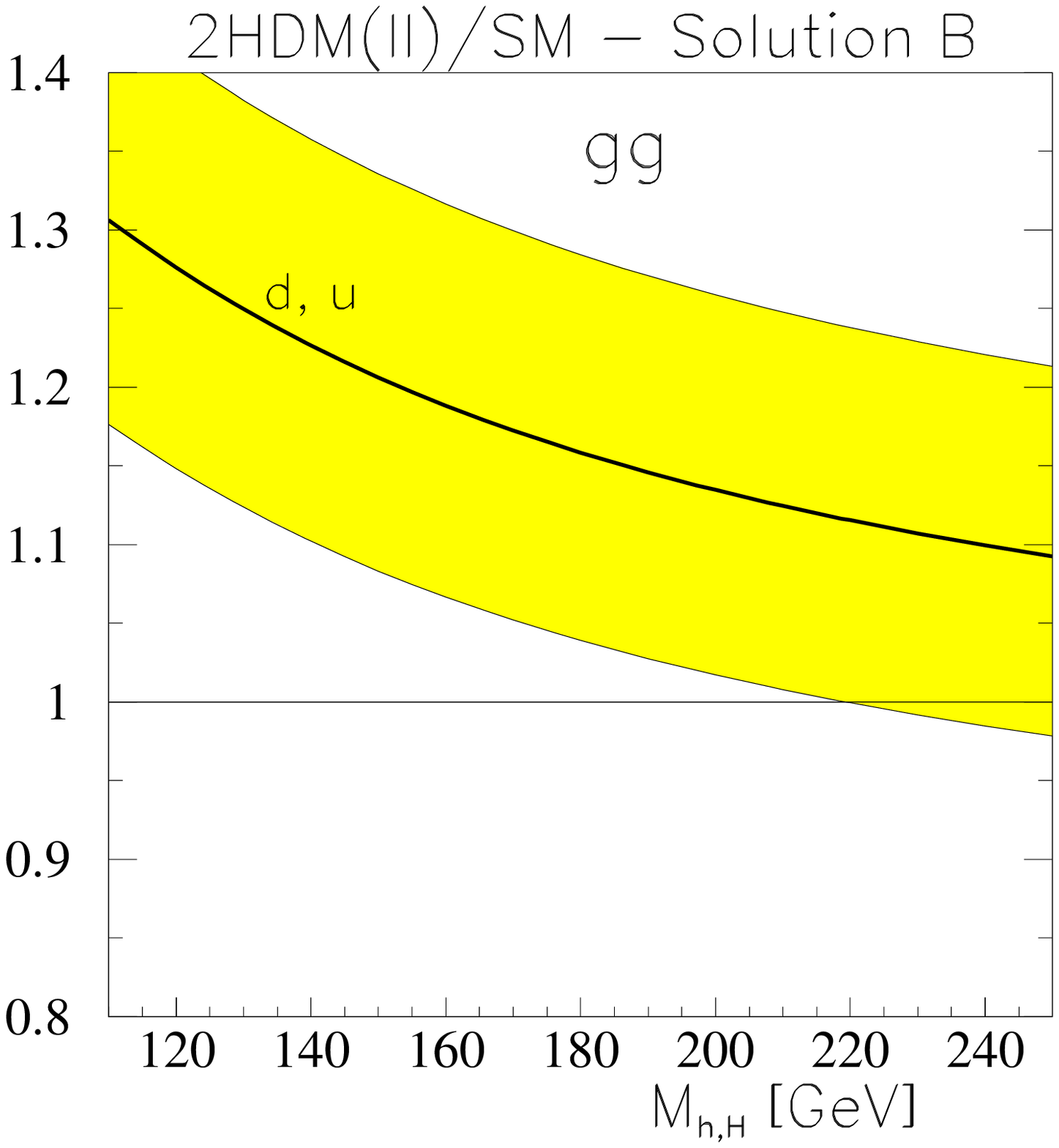}
\includegraphics*[scale=0.22]{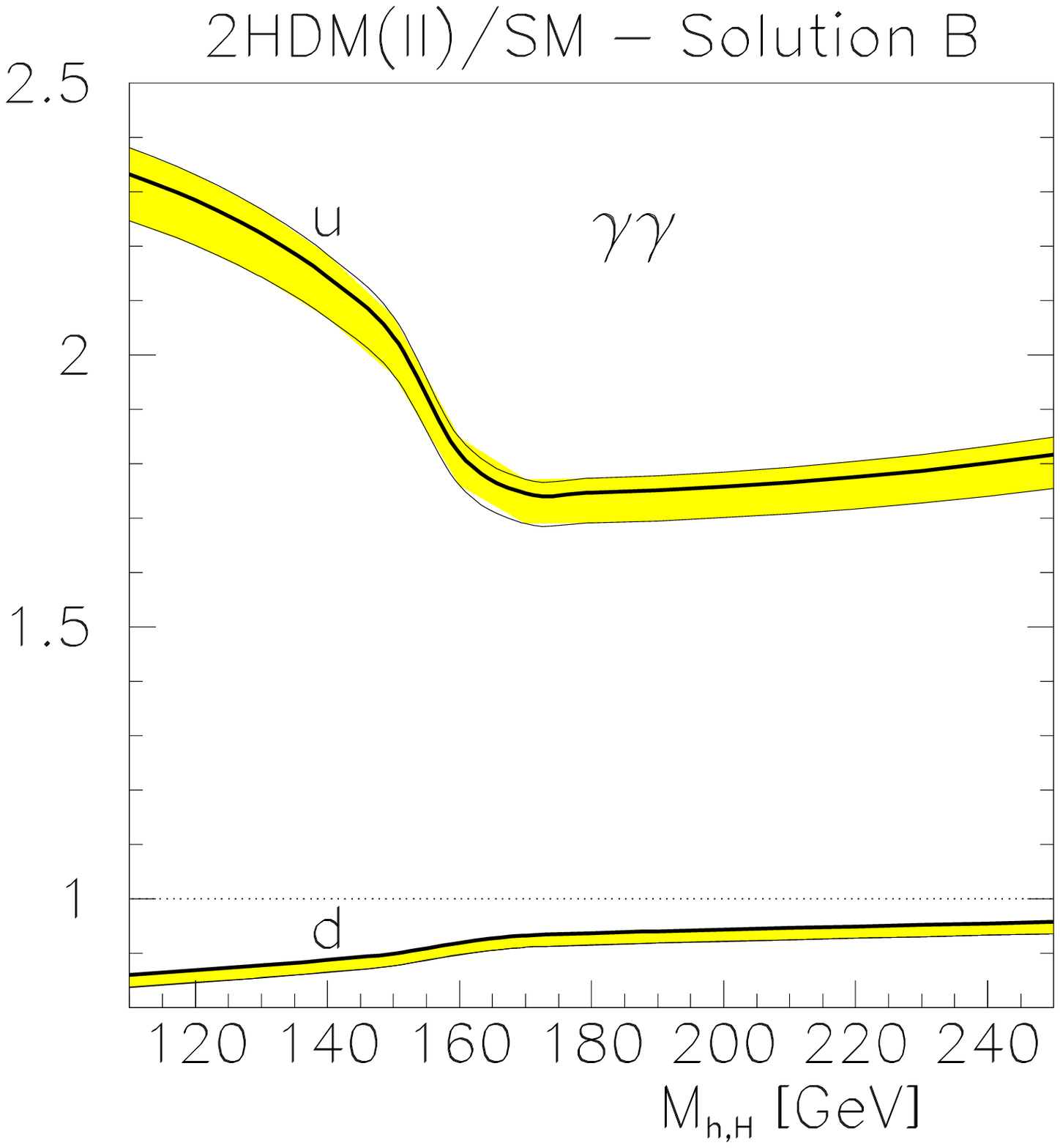}
\includegraphics*[scale=0.22]{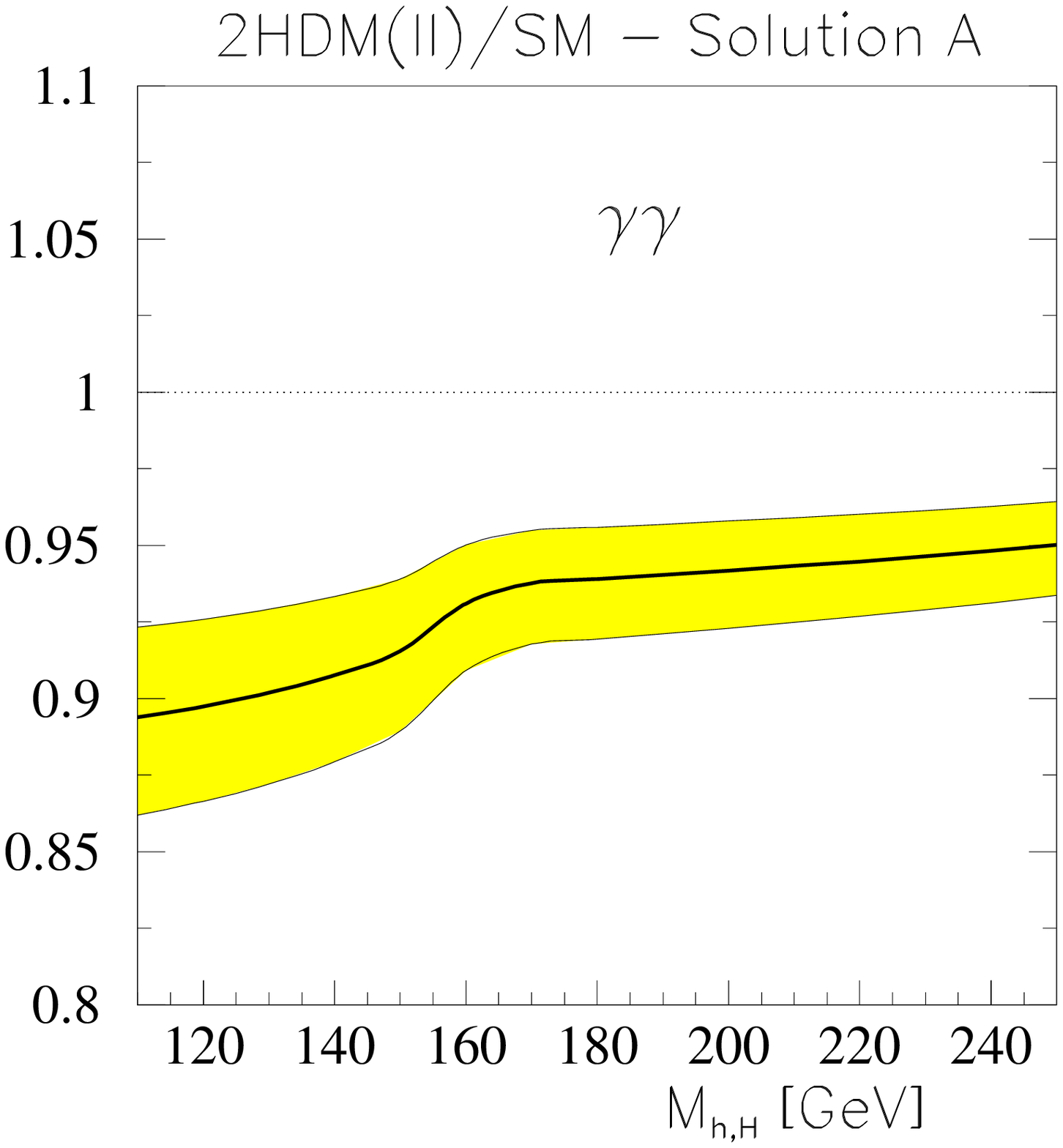}
\caption{The discrimination between a 2HDM and the SM using the 
$\gamma \gamma$ width at the PLC\protect\cite{lhc-ilc-2hdm}.\label{fig5}}
\end{figure}
In case of Model A, considered by the authors,
the $gg$ width is the same as that
in the SM whereas in Model B it differs by about $30 \%$--$40 \%$ from it. The 
first panel in Fig.~\ref{fig5} shows the $gg$ width in Model B, whereas the 
other two panels show the expected $\gamma \gamma$ widths for the two cases in 
comparison with the SM. One sees clearly that the accurate measurement of 
$\Gamma_{\gamma \gamma}$ possible at the PLC will indeed supplement the LHC data
substantially towards getting a more complete understanding of the 
spontaneous symmetry breaking.

\begin{figure}[htb]
\centerline{
           \psfig{figure=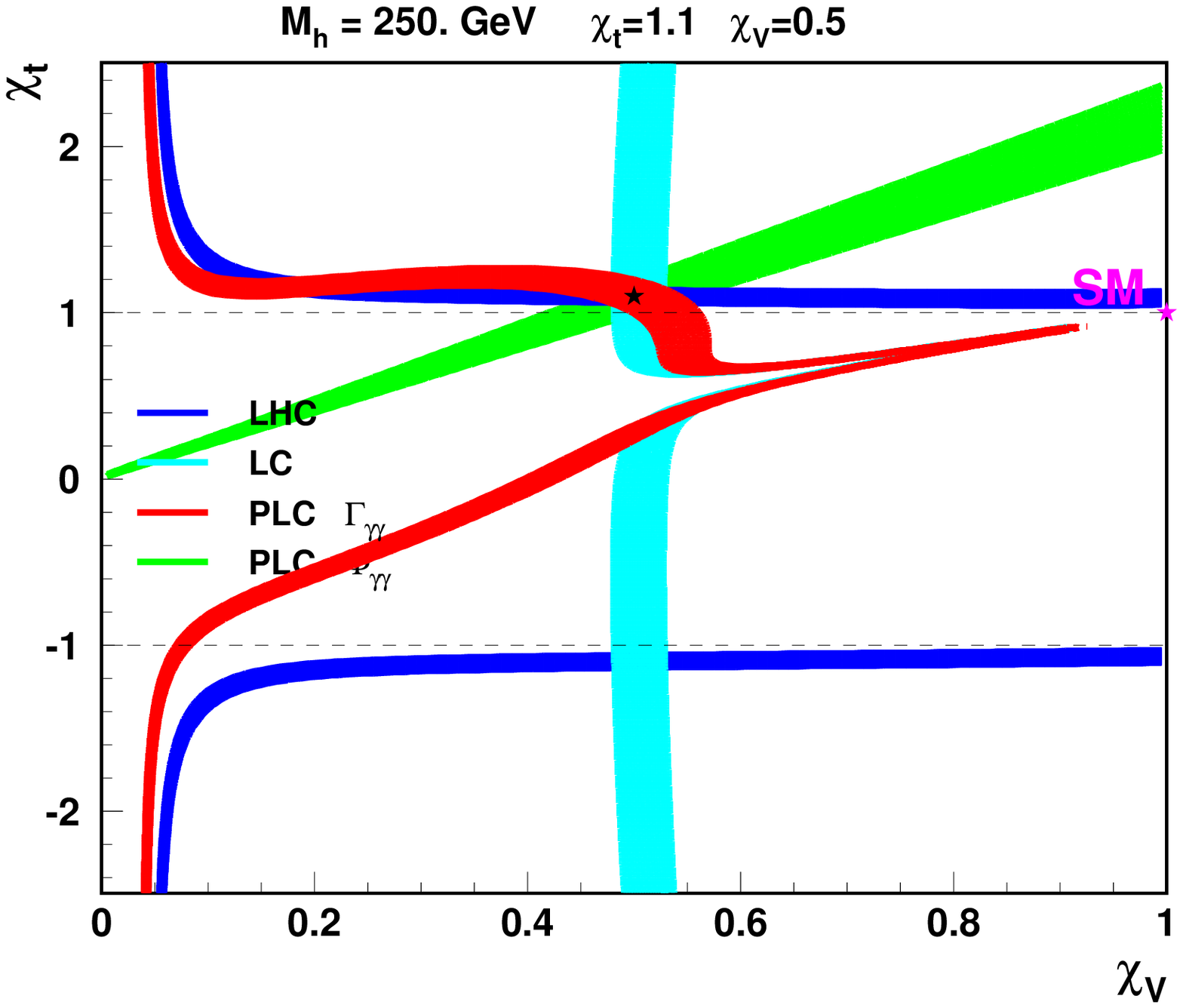,width=6cm,height=6cm,angle=0}
           \psfig{figure=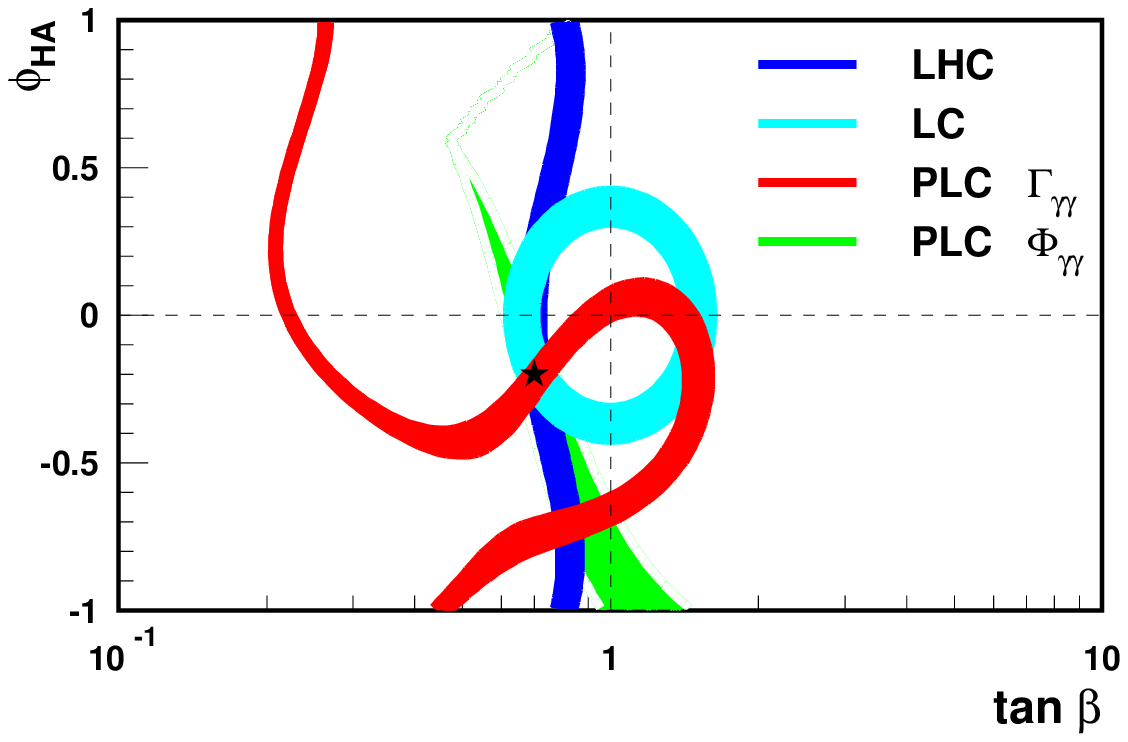,width=6cm,height=6cm,angle=0}}
\caption{The left panel shows the synergy between the LHC/ILC and the PLC in 
precise determination of the Higgs Boson couplings in a general 2 Higgs Doublet 
Model~\protect\cite{lhc-ilc-2hdm1}. The star at $(1,1)$ indicates the SM point 
and the 'star' at the centre of the plot corresponds to a particular set of
parameters for the general 2HDM for which the light Higgs has a SM-like 
phenomenology. The plot in the right panel shows an example of the same 
synergy for the case 
of he Higgs Boson couplings in a CP violating 2HDM~\protect\cite{lhc-ilc-cpv}.
The plot shows that the measurements at all the three colliders will be needed 
to determine conclusively whether the CP-violating phase is nonzero.  
\label{fig6}}
\end{figure}
For a heavier Higgs which can decay to a pair of gauge bosons, it is possible
at the PLC to measure the phase of the $H \rightarrow \gamma \gamma$  amplitude
through interference effects.
This phase carries information about the couplings of the H to the gauge bosons
as well as to a $t \bar t$ pair. On the other hand the LHC measurements can 
give better information on  $t \bar t H$ coupling whereas the ILC ones on
the  $h VV$ couplings. Hence combining the information from the PLC along
with the LHC and the ILC measurements, the couplings of a Heavy Higgs boson
can be pinned down too. This is illustrated in the left panel of Fig. 6 taken 
from 
Ref.~\cite{lhc-ilc-2hdm1}. One sees clearly that the $\phi_{\gamma \gamma}$
measurement possible at the  PLC will play an absolutely essential role in 
lifting the sign ambiguity which can not be resolved using the LHC and the ILC.
In case of CP-violating 2HDM, the  CP-violating phase will affect the 
phase of the amplitudes $H\gamma\gamma$ and $HWW$, which can be measured
via interference effects in the angular distributions of the decay $W$'s.
The plot in the right panel of Fig~\ref{fig6} shows clearly again the crucial 
role that the PLC can play in removing the ambiguities in the determination of 
the mixing angle in the $H$--$A$ sector.

Some more aspects of this synergy in the context of SUSY have been discussed 
in the proceedings elsewhere~\cite{mesusy}.
\subsection{Anomalous couplings and extra dimensions}
\begin{figure}
\includegraphics*[scale=0.3]{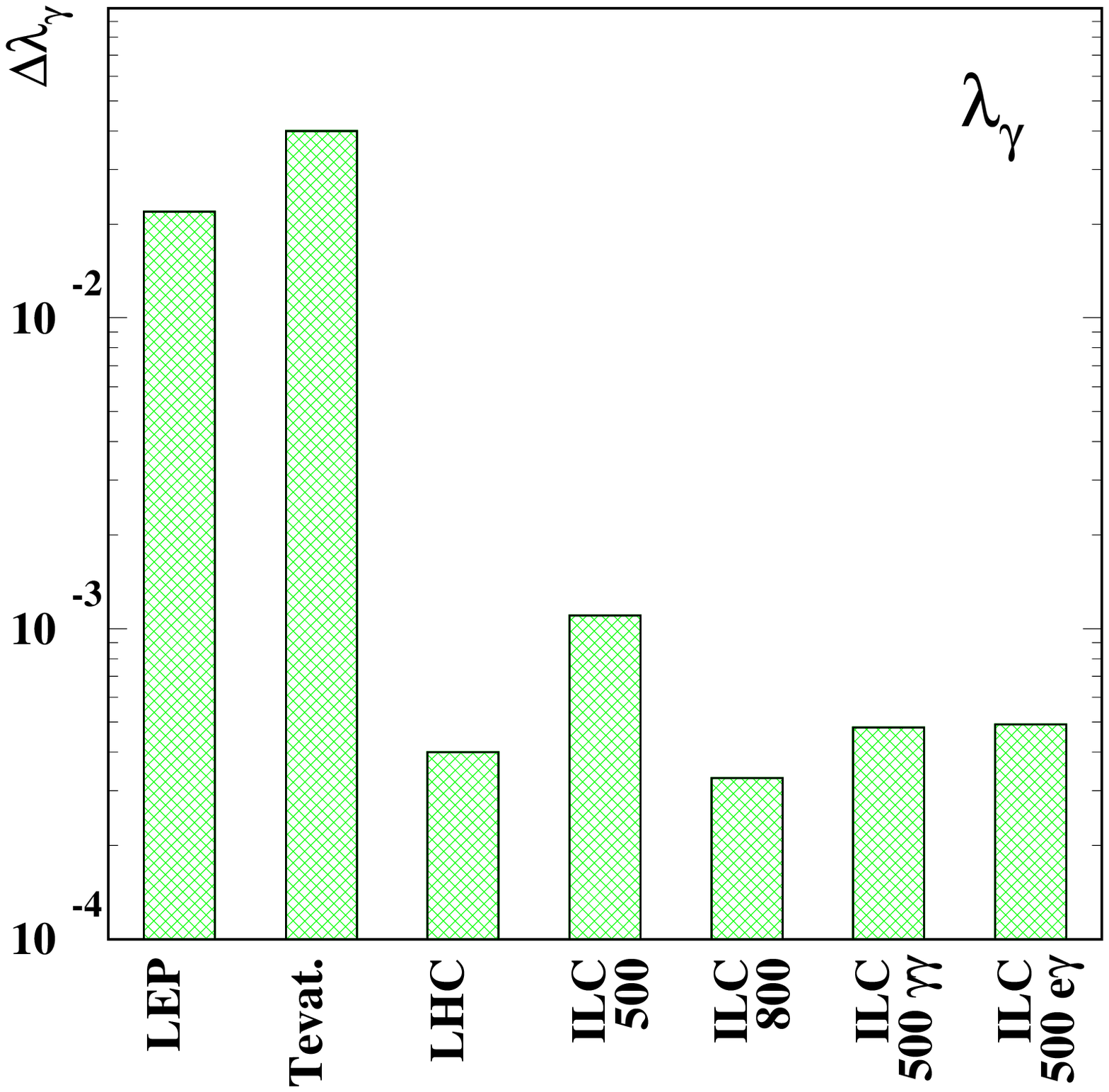}
\includegraphics*[scale=0.3]{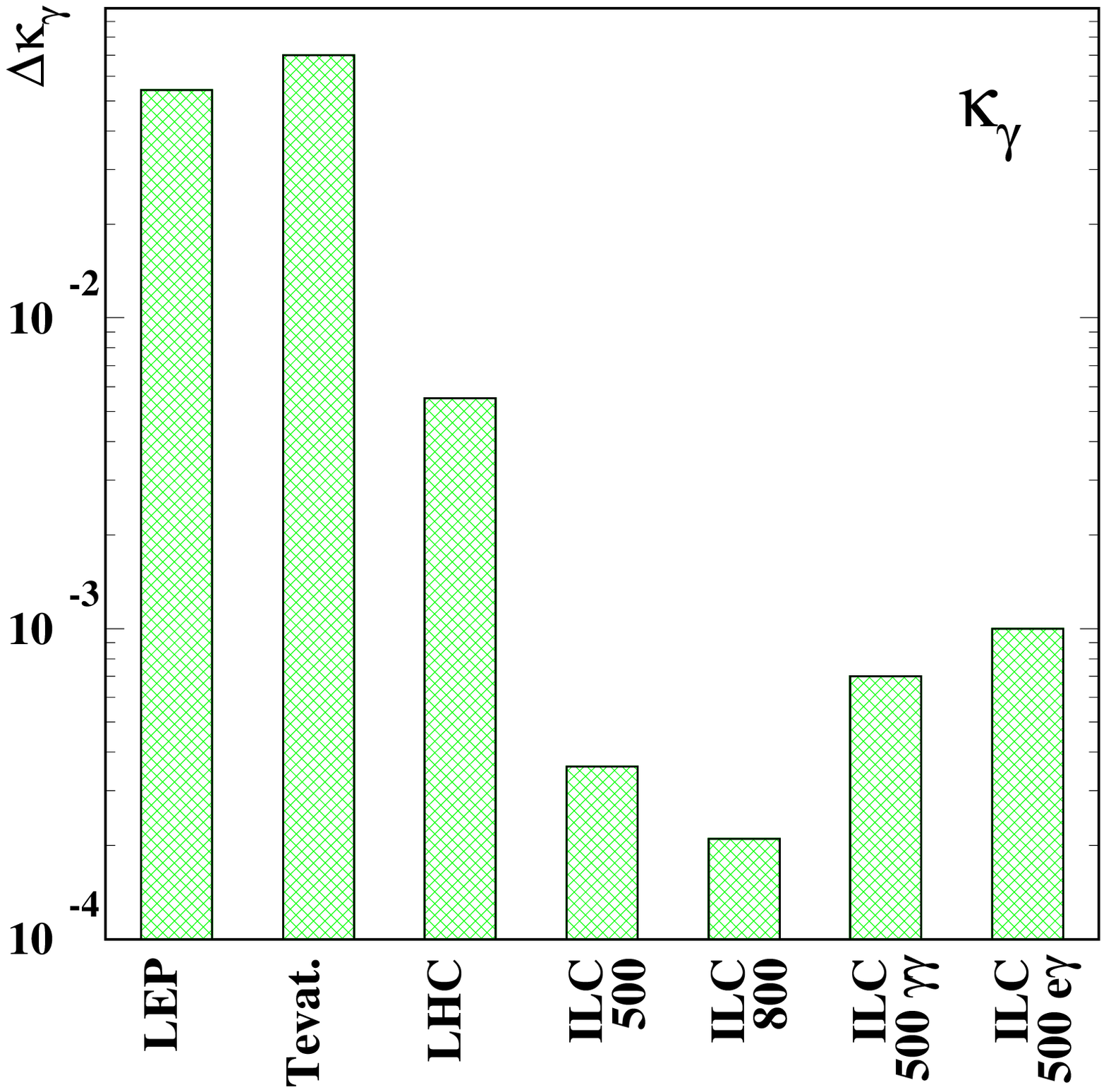}
\caption{A compilation of the reach of various colliders for the 
anomalous couplings of the photon~\cite{klaus}.\label{fig7}}
\end{figure}

One of the simplest ways to look for new physics related to EW symmetry 
breaking beyond  is through measurements of the anomalous couplings of the 
gauge bosons~\cite{tesla-tdr}, due to the contribution of the $t$-channel
diagrams. Fig.~\ref{fig7}~\cite{klaus} shows a comparison of the
potential of the different colliders for measurements of these anomalous 
couplings for the photon.
We see that while for the anomalous coupling $\lambda_g$ the PLC 
would perform better than a 500 GeV ILC, for the case of $\kappa_g$
the situation is different. This is a somewhat representative in this
context.

The PLC has interesting possibilities for the models with
extra dimensions since gravitons have large couplings to gluons
and photons, the polarisation of photons can also be used for spin
determination. Studies in the context of a $\gamma \gamma$ collider
do exist~\cite{tesla-plc,Ackermann:2004ag}. However, to my mind much more 
detailed analysis needs to be done in this case. One of the examples 
of things that still need to be done is discussed below.  For example, 
$t \bar t$ 
production in the process $\gamma \gamma \rightarrow t \bar t$  can be used 
very effectively~\cite{Mathews:1999ik} to probe the large extra dimensions.  
receives the usual QED $t/u$ channel contribution and the s-channel exchange 
of the tower of virtual Kaluza Klein particles. The 
ADD~\cite{Antoniadis:1998ig} model has two parameters: effective string scale 
$M_s$ and effective coupling $\lambda$ upto a sign ambiguity. The reach in 
$M_s$ can be already quite large just using rates. For a 500 GeV machine, 
the reach, eg., is 1.6 TeV. This can be further maximised by use of rapidity 
distributions and polarisation. However, this and similar other analysis are 
all done using the ideal backscattered laser spectrum~\cite{ginzburg}, which
however gets modified due to multiple interaction effects which have been 
calculated in a Monte Carlo simulation~\cite{telnov}. A convenient 
parametrisation of the realistic spectrum is available~\cite{zarnecki}.
The more realistic spectrum has a peak at low energies of the photon and 
the flux of the hard photon also gets by about a factor 2.
I present the update of the analysis~\cite{Mathews:1999ik}, using the 
more realistic spectrum~\cite{telnov,zarnecki}.
\begin{figure}
\includegraphics*[scale=0.25]{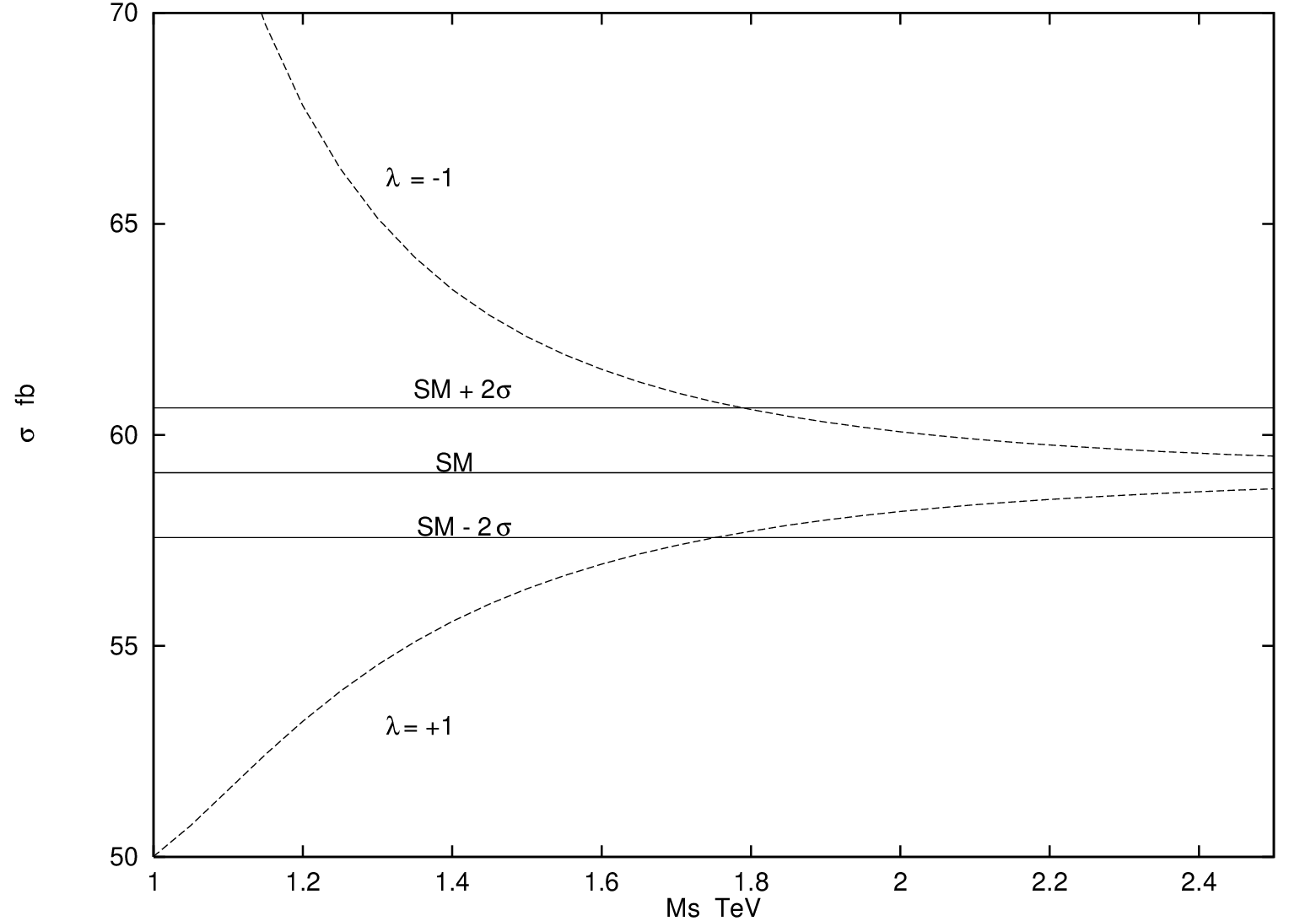}
\includegraphics*[scale=0.25]{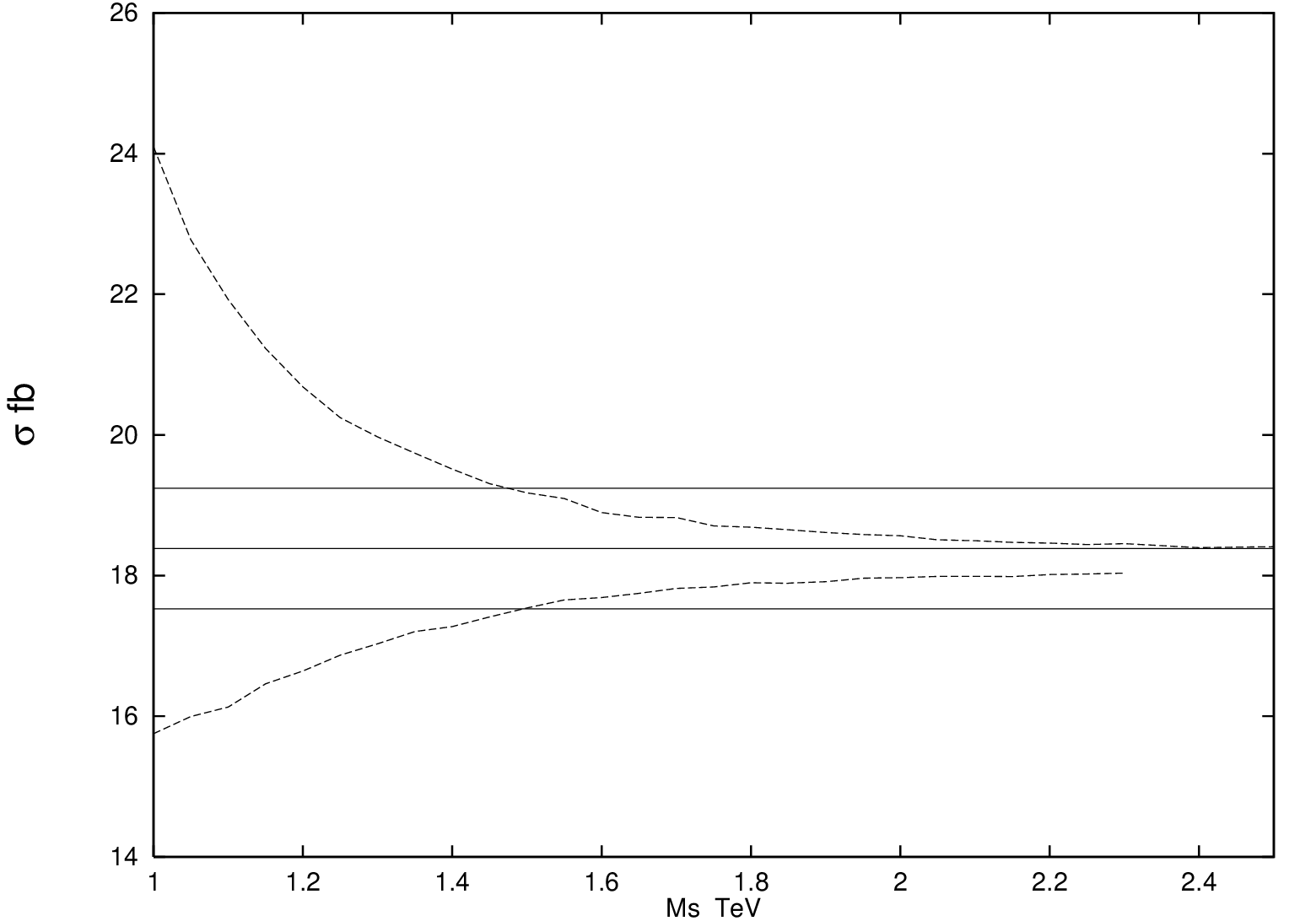}
\caption{Comparison of the limits obtained using the ideal backscattered
photon spectrum~\cite{ginzburg} (left panel) and the 
realistic ComPAZ~\protect\cite{zarnecki} (right panel).~\label{fig8}}
\end{figure}
Thus we see that this realistic spectrum does affect the limit substantially.
We can see from the right panel that the  sensitivity goes down from 1.7 TeV 
to 1.3 TeV, with the  realistic spectrum.

\section{Conclusions}
It is clear from the above short discussions that there is a large potential
for the LHC-ILC synergy and the study group document~\cite{lhc-ilc} has 
scratched only the surface so far. However, various good examples of the 
synergy have bee established quantitatively. There are certainly more 
ideas waiting to be thought about and studied. It is hard to believe, 
after these studies, that after ILC turn on no new questions will be asked of 
the LHC. It seems also clear that some overlap between LHC/ILC will therefore
be necessary. More work still necessary in the context of PLC, particularly
the use of its unique abilities in context of Extra Dimensional models.

\section{Acknowledgments}
It is a pleasure to thank the Organisers for the wonderful organisation
and the atmosphere at the meeting.

\end{document}